# Gendered Words and Grant Rates: A Textual Analysis of Disparate Outcomes in the Patent System

Deborah Gerhardt, Miriam Marcowitz-Bitton, W. Michael Schuster, Avshalom Elmalech, Omri Suissa & Moshe Mash[1]

Text is a vehicle to convey information that reflects the writer's linguistic style and communicative patterns. By studying these attributes, we can discover latent insights about the author and their underlying message. This article uses such an approach to better understand patent applications and their inventors.

While prior research focuses on patent metadata (i.e., filing year or gendered inventor names), we employ machine learning and natural language processing to extract hidden information from the words in patent applications. Through these methods, we find that inventor gender can often be identified from textual attributes—even without knowing the inventor's name. This ability to discern gender through text suggests that anonymized patent examination—often proposed as a solution to mitigate disparities in patent grant rates—may not fully address gendered outcomes in securing a patent.

Our study also investigates whether objective features of a patent application can predict if it will be granted. Using a classifier algorithm, we correctly predicted whether a patent was granted over 60% of the time. Further analysis emphasized that writing style—like vocabulary and sentence complexity—disproportionately influenced grant predictions relative to other attributes such as inventor gender and subject matter keywords.

Lastly, we examine whether women disproportionately invent in technological areas with higher rejection rates. Using a clustering algorithm, applications were allocated into groups with related subject matter. We found that 85% of female-dominated clusters (over 50% women inventors) have abnormally high rejection rates, compared to only 45% for male-dominated groupings.

These findings highlight complex interactions between textual choices, gender, and success in securing a patent. They also raise questions about whether current proposals (e.g., anonymized examination) will be sufficient to achieve gender equity and efficiency in the patent system.

---

[1] Deborah R. Gerhardt is the Paul B. Eaton Distinguished Professor of Law at the University of North Carolina School of Law.
Miriam Marcowitz Bitton is the Raoul Wallenberg Human Rights Full Professor of Law.
Mike Schuster is an associate professor at the University of Georgia's Terry College of Business with a courtesy appointment at the University of Georgia School of Law.
Avshalom Elmalech is an associate professor of Information Science at Bar Ilan University.
Omri Suissa holds a PhD in artificial intelligence and natural language processing from Bar Ilan University. He is currently a research scientist at Brown University and a co-founder and VP of R&D at ClearMash Solutions.
Moshe Mash holds a PhD in artificial intelligence from Ben-Gurion University. He is the founder of MashInnovateAI.

# Gendered Words and Grant Rates: A Textual Analysis of Disparate Outcomes in the Patent System

Deborah Gerhardt, Miriam Marcowitz-Bitton, W. Michael Schuster, Avshalom Elmalech, Omri Suissa & Moshe Mash

## I. Introduction

The intersection of gender and patenting has been the subject of significant empirical study. For example, female inventors file fewer applications, have their applications granted at lower rates, and secure patents of narrower breadth. There is, however, an important limitation in this line of literature. Most research analyzes patent metadata—such as inventor team size, technological field, and filing year. We deviate from this approach by employing computer science tools to explore the actual language used in patent applications. Through this methodology, we identify previously unreported gendered patterns in the patent ecosystem.

Our study addresses three primary questions using natural language processing and machine learning. First, is it possible to identify an inventor's gender based on the language used in an application? Second, can we predict whether a patent will be granted based on objective analysis, and if so, how important is inventor gender to this prediction? Third, after allocating applications into "clusters" with related subject matter, do male and female dominated clusters have different grant rates? Analysis supports each of these propositions.

To predict inventor gender from application text, we use a machine learning model to "classify" filings as coming from a male or female inventor. The model revealed that—even without knowing the inventor's name—it is possible to predict their gender through textual analysis. Such a finding raises important questions about textual choices as a function of inventor gender.

This conclusion is further honed through a corpus linguistics analysis. Beyond predicting gender, our model identifies specific words that disproportionately facilitate gender identification. This list is agnostic as to whether these terms identify a male or female inventor; they are just disproportionately likely to lead to a *correct* prediction. However, by comparing those words to a vast body of English text (i.e., a corpus linguistics analysis), we find evidence that these gender-identifying terms disproportionately identify female inventors. In application, it may be easier to distinguish female inventors from textual cues than their male counterparts.

We further investigate word choice and gender by analyzing whether these attributes influence application grant rate. Using a classifier algorithm, we can predict application success approximately two-thirds of the time. But perhaps more importantly, the algorithm can identify attributes that are particularly important to predicting if a patent will be granted. Interestingly, we again find that textual choices (e.g., measures of vocabulary diversity, word length, and sentence length) come to the fore. These attributes are the dominant characteristics in predicting patent grant, followed by inventor gender, and then subject matter keywords.

Building from this finding, we investigate whether men and women invent in different fields and whether those fields have disparate grant rates. To this end, we generate clusters of applications that claim related subject matter and identify clusters with disproportionately high rates of male or female inventors. Likewise, clusters with high rates of rejection are pinpointed. Analysis finds that more than 85% of female-dominated clusters have abnormally high rejection rates. Again, this raises questions about the interaction of grant rate, gender, and the field of invention.

This article proceeds in six parts. In Part I, we provide a brief overview of the patent system and then discuss gender disparities found in prior research. Part II describes the data collected and preprocessing activities undertaken to prepare for our text-based analysis. Part III sets forth descriptive statistics, and Part IV presents our working hypotheses and research methods. Part V sets forth our empirical findings arising from machine learning and natural language processing of patent applications. Part VI describes the importance of our findings to legal policy and suggests paths for future research. Included in Part VI is a corpus linguistics analysis that extends to the conclusions from our natural language processing by analyzing textual choices as they compare to their usage in everyday English.

## II. Background

Every year, the United States Patent and Trademark Office (USPTO) receives over 600,000 patent applications.[2] The vast majority are prosecuted by a patent agent or attorney.[3] After filing, the USPTO assigns a patent examiner with subject matter expertise to evaluate whether the claimed invention is patentable. If multiple statutory requirements are satisfied,[4] a patent will be granted.

Although women constitute more than half of the U.S. population, they represented only 12.8% of inventors who were granted a patent in 2019.[5] In this study, we aim to leverage machine learning techniques to gain latent insights into the factors contributing to such disparities.

Beyond simple inventor counts, gendered differences in patenting outcomes have been researched in several studies. Jensen, Kovács, and Sorenson analyzed 2.7 million US patent applications and concluded that women applicants succeeded in securing patents less frequently than men.[6] Schuster, Davis, Schley, and Ravenscraft examined 3.9 million applications and reached similar conclusions.[7] Both studies found evidence that the ability to identify an inventor's gender from their name might contribute to gender-disparate outcomes.

Taking these findings as a starting point, we embark on the task of addressing whether variables beyond gender-specific names contribute to this gap. To explore that question, we analyze not only the metadata of patent applications (e.g., filing year, subject matter, inventor attributes, etc.) but the text contained in patent applications.

Machine learning (ML) and natural language processing (NLP) are pivotal in developing algorithms that can analyze textual data to uncover latent patterns and insights.[8] For example, these tools have been used to identify the gender of a writer[9] or to identify a single author of multiple texts.[10] By leveraging large datasets and advanced techniques like deep learning, researchers can detect subtle linguistic and stylistic features that differentiate discrete groups of writers. For instance, ML models can analyze word choice,[11] syntactic structures,[12] and even punctuation patterns to predict a writer's gender[13] with considerable accuracy. We applied these tools to predict the inventor gender by analyzing the text of patent applications, and in doing so, we enhance our understanding of inventorship patterns and outcomes. Our methodology and results provide researchers with powerful new tools for more precise and insightful patent analysis.

---

[2] USPTO, https://www.uspto.gov/web/offices/ac/ido/oeip/taf/us_stat.htm (last accessed Aug. 8, 2024).
[3] Dennis Crouch, *Grant Rate by Size and Representation*, PATENTLY-O (Apr. 12, 2021, 2:59 PM), https://patentlyo.com/patent/2021/04/grant-rate-by-size-and-representation.html (finding that over 99% of non-provisional utility application filings involved representation by a U.S. patent practitioner in a sample of 30,000).
[4] *See, e.g.,* 35 USC 101–103.
[5] USPTO, *Progress and Potential: 2020 update on U.S. women inventor-patentees*, available online at https://www.uspto.gov/sites/default/files/documents/OCE-DH-Progress-Potential-2020.pdf.
[6] Kyle Jensen, Balázs Kovács & Olav Sorenson, *Gender Differences in Obtaining and Maintaining Patent Rights*, 36 NATURE BIOTECHNOLOGY 307, 307-09 (2018); *see also* W. Michael Schuster, R. Evan Davis, Kourtenay Schley & Julie Ravenscraft, *An Empirical Study of Patent Grant Rates as a Function of Race and Gender*, 57 AM. BUS. L. J. 281, 281–319 (2020) (same).
[7][7] Schuster, et al., *supra* note 6. This gender gap is not limited to the United States. Taehyun Jung & Olof Ejermo, *Demographic Patterns and Trends in Patenting: Gender, Age, and Education of Inventors*, 86 TECH. FORECASTING AND SOC. CHANGE 110, 110-24 (2014).
[8] Omri Suissa, Avshalom Elmalech, & Maayan Zhitomirsky-Geffet, *Text analysis using deep neural networks in digital humanities and information science*, 73 J. OF THE ASS'N FOR INFO. SCI. AND TECH. 268, 268–87 (2021).

---

[9] Francisco Rangel & Paolo Rosso, *Use of Language and Author Profiling: Identification of Gender and Age*, 177 NAT. LANG. PROCESSING & COGNITIVE SCI. 56, 56-66 (2013); Sravana Reddy & Kevin Knight, *Obfuscating Gender in Social Media Writing,* in PROCEEDINGS OF THE FIRST WORKSHOP ON NLP AND COMPUTATIONAL SOCIAL SCIENCE 17, 17-26 (Nov. 2016); Saman Daneshvar & Diana Inkpen, *Gender Identification in Twitter Using N-Grams and LSA*, in PROCEEDINGS OF THE NINTH INTERNATIONAL CONFERENCE OF THE CLEF ASSOCIATION (CLEF 2018) (Sept. 2018), available at https://ceur-ws.org/Vol-2125/paper_213.pdf; Babatunde Onikoyi, Nonso Nnamoko & Ioannis Korkontzelos, *Gender Prediction with Descriptive Textual Data Using a Machine Learning Approach*, 4 NAT. LANG. PROCESSING J. 100018 (2023).

[10] Moshe Koppel & Jonathan Schler, *Authorship Verification as a One-Class Classification Problem*, in PROCEEDINGS OF THE TWENTY-FIRST INTERNATIONAL CONFERENCE ON MACHINE LEARNING 62 (July 2004); Jacob Tyo, Bhuwan Dhingra & Zachary C. Lipton, *On the State of the Art in Authorship Attribution and Authorship Verification*, arXiv preprint arXiv:2209.06869 (2022); Mirco Kocher & Jacques Savoy, *A Simple and Efficient Algorithm for Authorship Verification*, 68 J. ASS'N FOR INFO. SCI. & TECH. 259-69 (2017); Siddharth Swain, Guarev Mishra & C. Sindhu, *Recent Approaches on Authorship Attribution Techniques—An Overview*, in 2017 INTERNATIONAL CONFERENCE OF ELECTRONICS, COMMUNICATION AND AEROSPACE TECHNOLOGY (ICECA), Vol. 1, 557-66 (Apr. 2017), IEEE.

[11] Felix Hamborg, Anastasia Zhukova, & Bela Gipp, *Automated Identification of Media Bias by Word Choice and Labeling in News Articles*, in 2019 ACM/IEEE JOINT CONFERENCE ON DIGITAL LIBRARIES (JCDL) 196-205 (June 2019), IEEE; Aparna Garimella, Rada Mihalcea & James Pennebaker, *Identifying Cross-Cultural Differences in Word Usage*, in PROCEEDINGS OF COLING 2016, THE 26TH INTERNATIONAL CONFERENCE ON COMPUTATIONAL LINGUISTICS: TECHNICAL PAPERS 674-83 (Dec. 2016).

[12] Juan Soler-Company, *Use of Discourse and Syntactic Features for Gender Identification*, NLP GROUP, POMPEU FABRA UNIVERSITY (2016); Rong Zheng, Jiexun Li, Hsinchun Chen & Zan Huang, *A Framework for Authorship Identification of Online Messages: Writing-Style Features and Classification Techniques*, 57 J. AM. SOC'Y FOR INFO. SCI. & TECH. 378-93 (2005).

[13] Efstathios Stamatatos, Nikos Fakotakis & George Kokkinakis, *Text Genre Detection Using Common Word Frequencies*, in COLING 2000 VOLUME 2: THE 18TH INTERNATIONAL CONFERENCE ON COMPUTATIONAL LINGUISTICS (2000); Amnah Alluqmani & Lior Shamir, *Writing Styles in Different Scientific Disciplines: A Data Science Approach*, 115 SCIENTOMETRICS 1071–85 (2018).

## III.  Data

To test our expectations regarding gender and textual choices in patent law, we amassed a dataset of approximately 270,000 published applications filed between 2013 and 2020, inclusive.[14] Each of these entries included the application number, the abstract, the filing year, and dummy variables for men inventors, women inventors, and applicants claiming "small entity" status.[15] We sorted the applications by technological field using the assigned USPC (United States Patent Classification) class and by allocating applications into one of 8 technological fields associated with the art unit it was assigned to at the USPTO.[16] Using USPTO data, we coded the outcome of each application as patented, abandoned, or pending as of June 22, 2022.[17]

The scope of our study was intentionally limited to remove potential noise that could muddy our efforts to identify trends associated with the inventor's gender. To focus on direct links between gender, patent attributes, and outcomes, we identified applications naming a single United States inventor. This allowed us to draw a direct link between inventor gender and relevant patent attributes and outcomes. By limiting the data to single inventor applications, we did not have to question whether one inventor might have caused a particular outcome or behavior—especially in mixed-gender teams. We recognize that limiting our dataset to single inventor teams may present certain biases (e.g., if single inventors tend to invent in certain areas or hire patent attorneys of a certain quality). However, we accepted this limitation to assure we were analyzing direct links between inventor gender and patenting outcomes.

We also limited our applications to U.S.-based inventors because due to cultural differences, foreign inventor behaviors may vary in a manner that is dissimilar to U.S. inventors. These possibilities could introduce noise into the associations we intend to study—namely patenting behaviors or outcomes that are disproportionately associated with inventors of a particular gender.

Our dataset includes only utility patent applications that did not claim priority to an earlier filing, except for priority claims to national stage entries or provisional applications. The choice to study only utility applications excluded design patents, plant patents, provisionals, reissues, etc. Given the prevalence of utility applications,[18] such a choice is common in the literature.[19] This decision

---

[14] This data included all entries for which an abstract could be identified and which satisfied all of the requirements outlined below. This led our dataset to include 39,128 applications from 2013, 38,961 applications from 2014, 36,574 applications from 2015, 34,937 applications from 2016, 33,553 applications from 2017, 32,729 applications from 2018, 31,198 applications from 2019, and 25,321 applications from 2020. Note that these are actual filing dates, not priority dates.

[15] 55.71% of the applications claimed small entity status.

[16] The eight technological classifications are associated with the art unit that an application was assigned to by the USPTO. The mutually exclusive classifications and the art unit they are associated with are as follows: Biotechnology (1600s), Chemical and Materials Engineering (1700s), Computer Architecture (2100s), Computer Networks (2400s), Communications (2600s), Semiconductors (2800s), Transportation (3600s), and Mechanical Engineering (3700s). These classifications are adopted from Schuster, et al., *supra* note 6, at 302 n.94.

The distribution of patents across the categorized technological fields is as follows: Biotechnology accounts for 5.89%, Chemical and Materials Engineering for 7.95%, Computer Architecture for 6.12%, Computer Networks for 6.82%, Communications for 8.29%, Semiconductors for 13.79%, Transportation for 25.38%, and Mechanical Engineering for 24.56% of all patents. No category was assigned for 1.22% of the applications.

[17] 62.04% of the applications were patented, 27.22% were abandoned, and 10.74% were pending.

[18] According to data from Patentsview, utility applications accounted for 92.7% of filings in 2020 (excluding provisional applications from the total).

[19] *See*, e.g, Jensen, et al., *supra* note 6, at 307.

furthered our goals of studying associations between gender and patent behaviors by excluding other types of filings that might have disparate outcomes by gender.

Patent applications may relate directly (i.e., make a claim of "priority") to previous filings relating to the same invention. We chose to remove applications making priority claims (other than claims to a provisional or foreign application) to reduce the possibility of double counting multiple applications associated with a single invention. Applications making priority claims to earlier filings will often list the same inventor and include the same abstract. If one patent family included 10 related applications, it might appear within our dataset 10 times—creating a disproportionate impact. We avoided this problem by keeping the first application in our dataset and removing those that followed. Applications claiming priority to a provisional or foreign application were not excluded because the original filings would not appear in our dataset, and therefore, double counting was not an issue.

After limiting our dataset to remove potential noise, we estimated whether inventors are men or women based on data from Gender API.[20] We coded inventor gender if the data identified it with at least a 70% certainty. Inventors who did not satisfy this threshold were coded with a zero for both Male and Female. Using this standard—out of 272,401 applications—234,100 inventors were coded as men, 26,718 were coded as women and 11,583 were not included in our analysis because we could not estimate their inventor's gender with sufficient certainty. In all, 95.7% of our applications were coded for gender and male inventors outnumbered female inventors by an 8.76 to 1 ratio. Additional descriptions of our data are available in **Appendix A**.

To further improve the quality and reliability of our analysis, we conducted additional cleaning activities.[21] First, we removed duplicate abstracts. While we attempted to achieve this goal by excluding applications making priority claims (other than claims to foreign filings or provisionals), we found a small percentage (2,489 of our 272,401 applications amounting to .91%) of repeated abstracts. Due to the low percentage of applications with repeated abstracts, we removed them to avoid double-counting.

Next, we sorted the applications into USPC classes. During the exploratory data analysis (EDA) process, we discovered that the column representing patent classes from the PatEx dataset contains 839 distinct values. According to the USPC website, there should be approximately 400 classes, indicating a potential anomaly in the data. Further analysis of the values within the column revealed that the class entries appear in mixed data types—specifically integers and strings. For instance, values related to class 72 were observed in two different formats: "072" and "72." To address this inconsistency, a custom function was implemented to convert all the class values to strings including leading zeros. After cleaning, 429 unique USPC classes remained.

As discussed above, approximately 4% of our applications were filed by inventors whose names did not meet the 70% threshold for identifying them as a man or woman. Due to the low rate of inventors who could not be coded for gender with sufficient certainty, we removed them from the dataset. We checked to see if any abstracts had multiple status values (accepted, rejected, pending), and found no outliers.

---

[20] GENDER-API, https://gender-api.com/ (last visited July 21, 2024). Gender API was recognized as a top performing gender-identification service in a benchmark study. Santamaría, L., & Mihaljević, H. (2018). Comparison and benchmark of name-to-gender inference services. PeerJ Computer Science, 4, e156. https://doi.org/10.7717/peerj-cs.15.

[21] Our code and data is found at: Moshik Mash, Patent Applicants Gender Classifier, GitHub (2024), https://github.com/MoshikMash/Patent_Applicants_Gender_Classifier.

We next pre-processed the textual data in each application's abstract to increase the quality and reliability of our results. Our team created a custom function to perform several additional steps. First, all text was converted to lowercase to ensure consistent formatting. Second, English stopwords, such as "the," "this" and "it"—which are commonly used but do not contribute substantive meaning[22]—were removed. Additionally, we eliminated numbers and punctuation marks. Last, the function applied lemmatization, which reduces words to their base or dictionary form.[23] We stored the results in a new column named "clean_abstract." As described below, we report results based on analysis of both original and clean abstracts.

To identify features associated with our cleaned abstracts, we used Python's Readability package.[24] Specifically, we employed the Flesch-Kincaid readability grade. This metric quantifies the difficulty in comprehending English passages based on the average number of syllables per word and the average number of words per sentence.[25] Its output estimates the grade level necessary to understand the text. A lower grade level score indicates easier readability, while a higher score suggests a higher level of complexity in terms of reading requirements.

Similarly, we added word-based statistics for each abstract. Again using Python's Readability package, we calculated the following for each abstract: word count, character count, median word length, average word length, word length asymmetry distribution degree (i.e., skewness), average characters per word, average words per sentence, number of syllables and sentences, type-token ratio (number of unique words divided by word count), number of long words (length over 7 characters), and number of complex words (more than 3 syllables).[26]

## IV. Descriptive Statistics

The applications in our data were filed between 2013 and 2020. After cleaning and pre-processing, the dataset contained 255,728 records with 37 columns. We divided the applications into 8 broader subject-matter categories by the USPTO-assigned examination group. The most dominant categories are Transportation (26.02%) and Mechanical Engineering (25.16%), which compose 51.18% of the data. The remaining categories are: Semiconductors (13.78%), Communications (8.27%), Chemical and Material Engineering (7.99%), Computer Networking (6.81%), Computer Architecture (6.12%), and Biotechnology (5.83%).

---

[22] W. Michael Schuster & Kristen Green Valentine, *An Empirical Analysis of Patent Citation Relevance and Applicant Strategy*, 59 AM. BUS. L. J. 231, 238 (2022) (Stop words include "nonsubstantive words such as 'you,' 'because,' and 'will.'").

[23] Quentin J. Ullrich, *Corpora in the Courts: Using Textual Data to Gauge Genericness and Trademark Validity*, 108 TRADEMARK REP. 989, 1038 n.280 (2018) ("Lemmatization is the process by which forms of a word are identified as instances of an underlying basic form. In corpus linguistics, these instances are frequently referred to as 'tokens,' and the underlying types are frequently referred to as 'types.'") (citing Christian Lehmann, *Lemmatization*, Linguistic Methodology (2013), http://www.christianlehmann.eu/ling/ling_meth/ling_description/lexicography/lemmatization.html.).

[24] *Python Readability Project*, https://pypi.org/project/readability/ (last visited Dec. 11, 2024).

[25] Tim R Samples et. al., *TL;DR: The Law and Linguistics of Social Platform Terms-of-Use*, 39 BERKELEY TECH. L.J. 47, 85 (2024) ("F-K score results range on a scale of zero to eighteen, which approximates the years of education required to understand the text. Thus, the lower the score, the more readable the text is.").

[26] We also introduced three additional columns to the dataset. These columns were created to convert the values in the gender, status, and parent category columns from a one-hot encoding format into a categorical format. This transformation enables easier analysis and prediction of the applicant's gender and the patent's status.

As noted above, the gender representation is highly imbalanced. Within our dataset, the Man:Woman ratio remained at 8.76:1 after cleaning. The ratio varied across each of our eight technological categories. As shown in Figure 1, Mechanical Engineering constituted the largest percentage of applications from women inventors; Transportation was the highest percentage for male inventors. Biotechnology is a dominant category for female applications (when comparing the percent of applications by gender by field), while Semiconductors is a dominant category for male applications. These trends are consistent with prior research.[27]

**Figure 1: Percentage of Filings in Technological Categories by Gender**

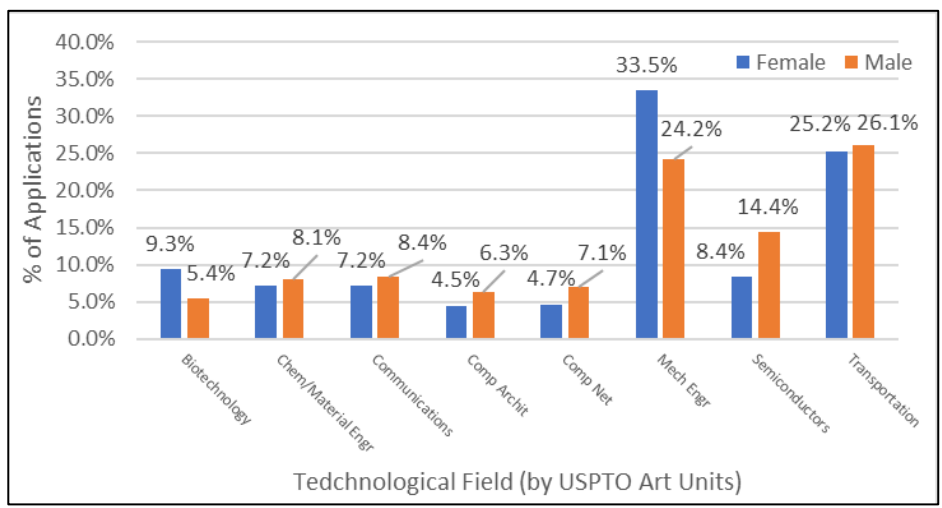

As shown in Figure 2, many textual characteristics of patent applications filed by men and women are similar. Each 'box' in Figure 2 represents the interquartile range (25% of the data above the median and 25% of the data below the median) of a particular textual metric with regard to men and women inventors. The number inside the box represents the median measurement for that metric for that group of inventors. All of the below findings present statistically significant differences by gender (as indicated by the p-values at the top right of each graphic).

Despite overlap of many of the below metrics, interesting gender differences exist in several text-based metrics. Applications filed by men have higher medians in words per sentence, complex words, and long words—suggesting that compared to women, men file patent applications with longer abstracts, more words per sentence, more syllables, and more complex words. This suggests that male and female inventors vary in *how* they describe their inventions—a theme that we will return to later.

---

[27] W. Keith Robinson, *Artificial Intelligence and Access to the Patent System*, 21 NEV. L.J. 729, 758 (2021) (describing female patent examiners being more likely to work in chemical and biological fields); Ann Bartow, *Patent Law, Copyright Law, and the Girl Germs Effect*, 90 ST. JOHN'S L. REV. 579, 593 (2016) (describing female STEM participants being more likely to work in biological fields).

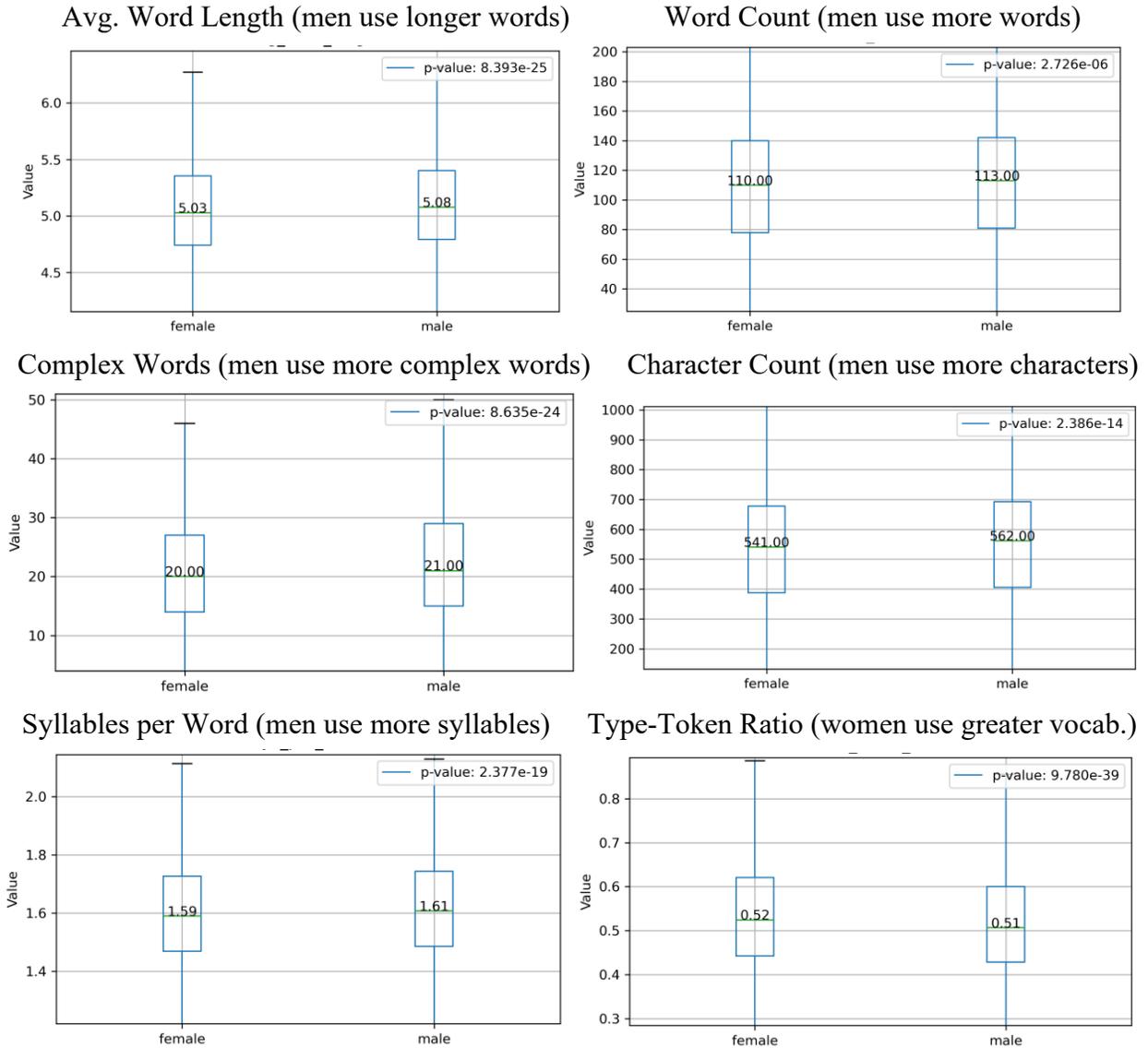

Figure 2: Gender-Based Analysis of Textual Characteristics

A full set of graphs depicting our analysis of gender-based textual characteristics is in **Appendix C**.

Within our dataset, a substantially greater percentage of male applications were granted (64.0%) versus female applications (52.0%).[28] All 8 of our technological classifications reflect this gender disparity.

## V. Hypotheses and Research Methods

After creating our dataset and conducting descriptive analysis, we ran tests to determine whether textual analysis can teach us more about the gender disparity in patent applications. To advance this goal, we tested three hypotheses. Due to the low absolute number of applications filed by women in several categories, we test our hypotheses within the Transportation and Mechanical

---

[28] The reject rate in female applications 37.0% is much higher than the male reject rate of 26.5%.

Engineering fields, which compose 51.18% of the data and have enough applications from women inventors to reflect statistically reliable results.

**Hypothesis 1**: It is possible to predict inventor gender based on the content of a patent abstract.

This hypothesis arises from different word use patterns in our abstracts as a function of gender. More specifically, preliminary analysis (shown below) found variations in the prevalent unigrams and bigrams within our abstracts. A unigram is a single token (e.g., a single word)[29] and a bigram is a set of two words in a particular order.[30] An "Ngram" is a string of terms (in a particular order) consisting of N words. The table below describes Ngrams that only appear in the top 50 Ngrams for male or female inventors (and not the other). Thus, an Ngram that appears in the top 50 for applications from men *and* women inventors will not appear on either list.

Table 1 lists 11 unigrams for women that did not appear in the top 50 most common unigrams for men (and vice-versa). Accordingly, 39 out of 50 most common unigrams for male and female inventors overlap. The table also lists 9 gender-specific bigrams for men and women (with an overlap of 41 out of 50 bigrams). As later described in Part VI, textual choices of this nature are germane to both gender-identification and whether a patent will be granted.

**Table 1: Gender-Unique Ngrams (Out of the Top 50 most Common for Each Gender)**

| Ngram Type | Only in Male Applications (out of 50 most common Ngrams) | | Only in Female Applications (out of 50 most common Ngrams) | |
|---|---|---|---|---|
| Unigram | Top | Comprises | Signal | Housing |
| | Panel | Opening | Control | Image |
| | Present | Cover | Vehicle | Associated |
| | Wherein | Attached | Power | Fluid |
| | Use | Container | Sensor | Component |
| | Comprising | | Housing | Communication |
| Bigram | Soybean variety | Main body | Soybean cultivar | Least partially |
| | Invention provides | One another | Plant part | Includes least |
| | Second side | Provided invention | Method may | First portion |
| | Also provided | Side wall | Light source | Control system |
| | Top surface | | Method apparatus | |

<u>Research Method</u>: To test this hypothesis, we train a classification model to predict the inventor's gender. After training on known data, the classifier analyzes an abstract and predict the inventor's gender. If the classifier achieves an accuracy score of above 50% the hypothesis is supported.

---

[29] Irene Nikkarinen, et al., *Modeling the Unigram Distribution*, in FINDINGS OF THE ASSOCIATION FOR COMPUTATIONAL LINGUISTICS: ACL-IJCNLP 2021, at 3721-29 (Assoc. for Computational Linguistics Aug. 2021)
[30] Dionysia Katelouzou, *The Rhetoric of Activist Shareholder Stewards*, 18 N.Y.U. J. L. & BUS. 665, 726 (2022).

We will use two classifiers—Logistic Regression based on the TFIDF matrix ("Term Frequency-Inverse Document Frequency")[31] and a DeBerta transformer model.[32] The classifier based on TFIDF is predominantly influenced by the specific words and the textual choices used in the text, whereas the DeBerta classifier demonstrates greater robustness, being more sensitive to the topics and underlying meanings within the abstracts. This distinction enables us to evaluate whether predictions are driven more by textual choices or the substantive topics of the abstracts.

**Hypothesis 2**: It is possible to predict the outcome of a patent application based on text characteristics and topics.

This hypothesis looks at the data from a different perspective. Exploratory analysis found that gender, category, content, and text characteristics correlate with an application's success. We therefore ask if it is possible to predict if a patent will be granted based on these features.

Research Method: To test this hypothesis, we will apply a Random Forest Classifier[33] to predict whether an application will be granted. If the Classifier's accuracy exceeds 50% (random guess), the hypothesis is supported.

The Random Forest Classifier can further identify features that particularly influence grant rates. Among others, we investigate textual choices and application subject matter. Textual choices are quantified using Python's Readability package to identify attributes like average sentence length and vocabulary usage. To extract subject matter keywords, an HDBSCAN clustering algorithm[34] was employed to group (i.e., "cluster") abstracts describing similar subject matter and identify keywords that are germane to those applications.

**Hypothesis 3**: Technological fields that are more common in female applications have a higher rejection rate.

Exploratory analysis found that female patent applications have a higher rejection rate, and that the men and women disproportionately invent in different technological fields. We thus ask whether male and female-dominated fields have different rates of rejection. Based on our exploratory analysis, we expect women inventors to work in areas with disproportionately low grant rates.

Research Method: This hypothesis is tested in two manners. First, we identify the top 25 USPC technological classes (assigned by the USPTO) for each gender and calculate each class's overall rejection rates. From this, we can assess whether female inventors file applications in classes with disproportionately rates of rejection. We next expand on this approach by using a clustering algorithm to break applications into groups that relate to the same subject matter. From there, it is possible to identify clusters that are predominately male or female and assess the grant rates of clusters as a function of dominant inventor gender.

---

[31] *See* Samah Senbel, *Fast and Memory-Efficient TFIDF Calculation for Text Analysis of Large Datasets*, in ADVANCES AND TRENDS IN ARTIFICIAL INTELLIGENCE: ARTIFICIAL INTELLIGENCE PRACTICES 557-563 (H. Fujita, A. Selamat, J. CW. Lin & M. Ali eds., 2021) (discussing term frequency – Inverse Document Frequency (TFIDF)).

[32] *See* P. He, X. Liu, J. Gao, & W. Chen, *DeBERTa: Decoding-enhanced BERT with Disentangled Attention*, arXiv e-prints, arXiv-2006 (2020).

[33] See Mahesh Pal, *Random Forest Classifier for Remote Sensing Classification*, 26 INT'L J. REMOTE SENSING 217 (2005).

[34] *See* Geoffrey Stewart & Mahmood Al-Khassaweneh, *An Implementation of the HDBSCAN* Clustering Algorithm*, 12 APPLIED SCI. 2405 (2022).

## VI. Data Analysis

This portion of our study focuses on testing the hypotheses we set forth above. Specifically, we address three primary questions. Can inventor gender be identified by textual content of an application? Second, can patent grant be predicted by a machine learning algorithm and what attributes of an application are particularly relevant to this prediction? And third, do men and women invent in different fields and do male or female-dominated areas have different grant rates.

### A. Hypothesis 1 – Identifying Inventor Gender from Textual Choices

We test Hypothesis 1 to see if inventor gender can be predicted from the abstract's textual content. To this end, two text classification models are used to analyze applications from our dominant fields: transportation and mechanical engineering. Conducting within-category evaluation allows the models to focus on nuances of word usage, as opposed to common terms in particular fields. Using this approach, we are able to predict the inventor's gender correctly 57–68% of the time.[35]

To prepare the abstracts for analysis, stop words (i.e., "the" or "or"[36]) were removed, as they provide little relevant information to the model. We then normalized different versions of the same word (e.g., singular and plural). This allowed us to homogenize related terms prior to analysis.

Our first model used a Logistic Regression Classifier. This approach is commonly employed for binary classifications like predicting inventor gender. To prepare for this analysis, we used a TFIDF Matrix ("Term Frequency-Inverse Document Frequency") to convert our textual data into a numerical format that can be processed.

More specifically, the TFIDF matrix transforms our abstracts into vectors representing the importance of all terms in our text. From there, logistic regression uses these vectors to identify patterns that are used in classifying the text as "male" or "female" inventor.

Our second classification model is based on DeBERTa ("Decoding-enhanced BERT with Disentangled Attention"). DeBERTa is a transformer-based model that analyzes complex language by capturing nuanced relationships within the text. Its advanced approach to natural language processing leverages self-attention mechanisms (i.e., understanding the relationships between words) to understand the context and semantics of an abstract's content.

Our DeBERTa model processed the abstracts through layers of self-attention and feed-forward neural networks to generate contextual embeddings. Then, using a classification head, embeddings were classified as female or male. These embeddings embody the meaning and context of each word in relation to the entire abstract, providing a richer representation for classification.

The initial step in using our classifiers is training. Training a classifier model teaches the algorithm to recognize patterns in data such that it can make accurate predictions on other abstracts (that weren't used in training). In our case, we trained classifiers to predict inventor gender, which we coded into numeric values: 0 for "Male" and 1 for "Female." Considering the highly imbalanced

---

[35] Textual analysis of a patent's abstract is common in the literature. *See,* e.g., Sam Arts, Bruno Cassiman & Juan Carlos Gomez, *Text Matching to Measure Patent Similarity*, 39 STRATEGIC MGMT. J. 62, 64-65 (2018); W. Michael Schuster & Kristen Green Valentine, *An Empirical Analysis of Patent Citation Relevance and Applicant Strategy*, 59 AM. BUS. L.J. 231, 252 (2022).

[36] Henry Dao et. al., *Cooling Regulation on A Heating Planet: A Sentiment Analysis of Public Comments on the Sec's Climate Disclosure Rule*, 19 RUTGERS BUS. L. REV. 1, 28 (2024)

nature of the dataset (about 9 male inventors for every female inventor), an under-sampling[37] approach was adopted. Under-sampling ensures that analysis is conducted on an equal number of applications from male and female inventors. Among other benefits, this approach prevents overfitting on the male class.

The classifiers were run nine times, each time using a different random sample of the majority class, matching the size of the minority class. The final accuracy of the classifier was determined by averaging the accuracy results from all iterations. During each iteration, the data was split into 80% for training and 20% for testing. Our results are reported in Table 2:

**Table 2: Predicting Inventor Gender**

| Category | Model | Accuracy |
| --- | --- | --- |
| Transportation | Logistic Regression | 65.9% |
| Mechanical Engineering | Logistic Regression | 68.6% |
| Transportation | DeBerta | 57.3% |
| Mechanical Engineering | DeBerta | 58.7% |

Our Logistic Regression classifier successfully predicted inventor gender 67.45% of the time. The DeBerta model had a slightly lower success rate of 58.36%. As both figures exceed 50%, Hypothesis 1 is supported. But while both models support our hypothesis, their disparate success rates may suggest underlying differences in abstracts as a function of inventor gender.

As mentioned above, the DeBERTa and Logistic Regression models approach their predictions from different angles. DeBERTa excels in understanding the abstract's meaning by capturing nuanced textual relationships and deeper semantic information. Accordingly, DeBERTa can interpret meaning in ways that go beyond word frequencies or term usage. For example, DeBERTa might associate abstracts describing certain types of inventions with men or women.

On the other hand, the Logistic Regression classifier focuses more on textual choices. By converting the text into a TFIDF matrix, Logistic Regression emphasizes individual terms and their direct contributions to the classification task. This approach is thus more attuned to specific words and their significance in the abstract. As an example, Logistic Regression might home on specific patterns in language use to identify inventor gender.

Accordingly, these models provide complementary insights. DeBERTa uncovers subtle cues about the abstract's meaning, while Logistic Regression highlights the impact of textual choices. And while both predicted inventor gender better than chance, the Logistic Regression achieved a significantly higher accuracy. This may suggest that textual choices (the province of Logistic Regression) are more gender-indicative than the underlying meaning of the abstract (where DeBerta excels).

Such a conclusion could raise important questions about whether patent applications from male or female inventors differ more in *how* they are written (e.g., textual choices) or in *what* they say (e.g., invention subject matter). Within our study, Logistic Regression's superior performance suggests that the gender difference in patent applications may be more about textual choice than underlying subject matter. This question is, of course, not conclusively answered here.

---

[37] Undersampling randomly removes entries from the larger category (here, male inventor applications) until the larger and smaller categories are the same size. Jonathan H. Choi, *An Empirical Study of Statutory Interpretation in Tax Law*, 95 N.Y.U. L. REV. 363, 441 (2020).

### B. Hypothesis 2 – Predicting Patent Grant

In this part, we investigate Hypothesis 2, which proposes that a classifier can predict whether a patent will be granted. To this end, our classifier was fed a variety of information about each application, including meta-data (e.g., inventor gender), information on textual choices (like vocabulary breadth or average sentence length), and keywords extracted from the abstract. After training, the algorithm can predict whether applications will be granted and identify what application attributes are disproportionately predictive of success.

Our classifier exhibited a 62% success rate, which supports our expectations. But perhaps more notable are the attributes that disproportionately predict that a patent will be granted. As identified at the end of the last section, our earlier results suggest that male and female inventors may be best identified by their *style* of writing (i.e., textual choices), as opposed to the subject of their invention. Interestingly, the current section finds that textual choices (like vocabulary scope and variations in word length) are particularly important in predicting whether a patent will be granted.

For the current analysis, we note that applications in our dataset were rejected approximately 40% of the time. Thus, if a classifier can achieve an accuracy score exceeding 50%, it would support our hypothesis (since we used an under-sampling method to balance our dataset, it contains 50% male and 50% female patents in every experiment). To test this expectation, we employed a Random Forest Classifier.[38]

Random Forest Classifiers use multiple decision trees to make predictions. A decision tree splits data into subsets based on different attributes, aiming to create groups as homogeneous (i.e., share the same features) as possible. Each tree in a Random Forest makes an independent prediction about an outcome (e.g., patent grant), and the classifier combines these predictions to form a final, more accurate decision. This approach leverages the simplicity and interpretability of decision trees while improving reliability through aggregation—making it well-suited for predicting whether a patent will be granted.

Beyond making predictions about whether an application will be granted, the Random Forest Classifier can identify specific factors that are important to its prediction. That makes it possible to compare the relative importance of textual choices (e.g., average word length), underlying subject matter (e.g., topical keywords or technological classifications), and metadata (e.g., filing year or inventor gender). This multifaceted approach enables a deeper understanding of the factors influencing patent grants and rejections.[39]

---

[38] *See* Joseph Simpson, *How Machine Learning and Social Media Are Expanding Access to Mental Health*, 2 GEO. L. TECH. REV. 137, 141 (2017); David Lehr & Paul Ohm, *Playing with the Data: What Legal Scholars Should Learn About Machine Learning*, 51 U.C. DAVIS L. REV. 653, 661 (2017).

[39] Further data processing was necessary prior to applying the Random Forest Classifier. We converted the USPC class column into a one-hot-encoded matrix (i.e., a vector with a binary representation of the USPC classes that a machine learning model can use as an input) and removed data superfluous to this portion of the study. Specifically, the columns 'application_number', 'one_if_male', 'one_if_patented', 'one_if_abandoned', 'one_if_pending', 'abstract', 'clean_abstract', 'Gender_Target', and 'Parent_Category' were removed from the dataset as they were deemed non-relevant for the classification task at hand.

However, before analysis, we had to associate each abstract with subject matter keywords that could be fed into the classifier. To this end, we used an HDBSCAN clustering algorithm[40] to identify clusters of applications related to the same topic. After allocating applications into clusters, the three-most common nouns from each cluster were identified. We then associated these "cluster keywords" with every patent in that cluster.[41] Providing the cluster keywords to our classifier algorithm allows us to consider application subject matter in predicting whether a patent will be granted.[42]

Again, we used an under-sampling method to even out the number of abstracts from each gender and allocated 80% of our data to training and 20% to analysis. After training, the classifier—with 500 estimators and a maximum depth of 150—correctly identified whether an application was granted 62% of the time.

Next, we applied the Random Forest Feature Importance function to identify the most important application attributes within the classifier. The following table presents the top 30 most significant features in predicting whether a patent will be granted and identifies each feature as pertaining to a Textual Choice (e.g., syllables per word), Technological Classification (e.g., USPC category), Cluster Keyword (e.g., "system" or "user"), or Metadata Attribute (e.g., inventor gender). We note that textual choices dominate the most important attributes in predicting patent grant.

---

[40] *See* Geoffrey Stewart & Mahmood Al-Khassaweneh, *An Implementation of the HDBSCAN\* Clustering Algorithm*, 12 APPLIED SCI. 2405 (2022).

[41] The cluster topics were identified in the following manner:
  1. **Sentence-level breakdown and sentence embedding**: We divided the abstracts into sentences. Subsequently, we created sentence embeddings using a DeBerta model.
  2. **Dimensionality reduction with UMAP**: To improve computational efficiency, we applied the UMAP algorithm to reduce the dimensionality of the data. By running this algorithm, the data was transformed from the initial 768-dimensional vector space to a 5-dimensional space.
  3. **Clustering with HDBSCAN**: For clustering, we employed the HDBSCAN algorithm, which utilizes density-based clustering. We chose this algorithm as it does not require predefining the number of clusters and can handle clusters of varying densities. After several attempts, we set the hyperparameter min_cluster_size to 7.
  4. **Topic selection:** From each cluster, we choose the two most center sentences that are the most representable sentences of that cluster (main topic). From each of the sentences we selected the most three common nouns from unigrams and bigrams as topics.

[42] To enable efficient processing, a version of Random Forest utilizing GPU memory was employed. cuML, RAPIDS, https://github.com/rapidsai/cuml (last visited Aug. 10, 2024).

## Table 3: Features that Impact Whether a Patent is Granted

| Feature | Importance (Descending, multiplied by 100) | Feature Type |
|---|---|---|
| Type-token ratio (vocabulary) | 4.17 | Textual Choice |
| Skew word length (variation in length) | 3.80 | Textual Choice |
| Syllables per word | 3.79 | Textual Choice |
| Average word length | 3.78 | Textual Choice |
| Characters per word | 3.76 | Textual Choice |
| Character count | 3.72 | Textual Choice |
| Syllables | 3.64 | Textual Choice |
| Words per sentence | 3.60 | Textual Choice |
| Word count | 3.54 | Textual Choice |
| Long words | 3.35 | Textual Choice |
| Complex words (3+ syllables) | 3.25 | Textual Choice |
| Sentences | 2.08 | Textual Choice |
| Sentences per paragraph | 2.07 | Textual Choice |
| Median word length | 1.63 | Textual Choice |
| Parent category | 0.98 | Technological Classification |
| System | 0.84 | Cluster Keyword |
| Assembly | 0.84 | Cluster Keyword |
| User | 0.82 | Cluster Keyword |
| Male inventor | 0.78 | Procedural Attribute |
| Device | 0.76 | Cluster Keyword |
| Method | 0.74 | Cluster Keyword |
| Invention | 0.72 | Cluster Keyword |
| Surface | 0.66 | Cluster Keyword |
| Portion | 0.66 | Cluster Keyword |
| Position | 0.64 | Cluster Keyword |
| End | 0.60 | Cluster Keyword |
| Disclosed | 0.60 | Cluster Keyword |
| Body | 0.57 | Cluster Keyword |
| Apparatus | 0.56 | Cluster Keyword |
| Embodiment | 0.56 | Cluster Keyword |
| Use | 0.56 | Cluster Keyword |

Analyzing the 30 most significant features revealed that the most influential attributes are characteristics associated with textual choices—including measures such as Type-Token Ratio (number of unique words divided by word count), average word length, and word count. Inventor gender fell outside the top 10 most influential aspects of a patent application, though it was more important than all but three cluster keywords.

These results (i.e., the dominance of Textual Choices in predicting patent grant) strengthen our prior observations about the importance of *how* a patent is written. Given the historical gender disparities in patent grant rate, our findings reinforce the need for further investigation into the differences in how (e.g., textual choices) applications submitted by men and women are presented to the USPTO.

Further information about the success rate of our classifier algorithm and additional comparisons between the textual choices in granted patents and rejected applications are available in **Appendix D**.

### C. Hypothesis 3 – Gender Differences in Technical Classes

Hypothesis 3 considers whether women disproportionately invent in technological classes with higher rejection rates. Our prior findings emphasized the importance of textual choices (style) over subject matter keywords (technological field) regarding gendered outcomes in patent law. It is thus prudent to examine whether and to what extent differences in inventive field by gender relate to success in patent prosecution.

This analysis proceeds in two parts. We look to gender differences across United States Patent Classification (USPC) invention categories and a series of ad hoc generated subject matter clusters. In both instances, we find evidence that female inventors work in fields with disproportionately high rejection rates.

We first calculate the top 25 USPC categories for women and men inventors and the likelihood of success in prosecuting patents in each class. For example, USPC Class 705 (dealing with data processing) is the most common class for all inventors, as shown in Tables 4 and 5 below. In each table, we color-coded the classes by rejection rate. Green classes have a rejection rate that is lower than 50%. Red classes have a rejection rate that is higher than 50%. Pending applications were not considered in the analysis. The following tables present the results:

Table 4: Top 25 Male USPC Categories

|   | USPC Class | Percentage of Apps. | Probable Outcome |
|---|---|---|---|
| 1 | 705 | 4.1% | Reject |
| 2 | 709 | 1.7% | Accepted |
| 3 | 701 | 1.6% | Accepted |
| 4 | 340 | 1.5% | Accepted |
| 5 | 370 | 1.5% | Accepted |
| 6 | 345 | 1.4% | Accepted |
| 7 | 800 | 1.4% | Accepted |
| 8 | 455 | 1.4% | Accepted |
| 9 | 707 | 1.3% | Accepted |
| 10 | 726 | 1.3% | Accepted |
| 11 | 424 | 1.3% | Reject |
| 12 | 600 | 1.3% | Accepted |
| 13 | 348 | 1.2% | Accepted |
| 14 | 606 | 1.2% | Accepted |
| 15 | 52 | 1.2% | Accepted |
| 16 | 362 | 1.1% | Accepted |
| 17 | 280 | 1.0% | Accepted |
| 18 | 248 | 1.0% | Accepted |
| 19 | 715 | 1.0% | Accepted |
| 20 | 514 | 1.0% | Accepted |
| 21 | 382 | 0.9% | Accepted |
| 22 | 73 | 0.9% | Accepted |
| 23 | 700 | 0.8% | Accepted |
| 24 | 166 | 0.8% | Accepted |
| 25 | 713 | 0.8% | Accepted |

Table 5: Top 25 Female USPC Categories

|   | USPC Class | Percentage of Apps. | Probable Outcome |
|---|---|---|---|
| 1 | 705 | 4.7% | reject |
| 2 | 2 | 3.6% | reject |
| 3 | 424 | 3.2% | reject |
| 4 | 514 | 1.8% | accepted |
| 5 | 119 | 1.8% | accepted |
| 6 | 5 | 1.8% | accepted |
| 7 | 132 | 1.7% | reject |
| 8 | 206 | 1.6% | accepted |
| 9 | 604 | 1.6% | accepted |
| 10 | 340 | 1.6% | accepted |
| 11 | 600 | 1.5% | accepted |
| 12 | 434 | 1.5% | reject |
| 13 | 800 | 1.5% | accepted |
| 14 | 220 | 1.3% | accepted |
| 15 | 435 | 1.3% | accepted |
| 16 | 224 | 1.2% | accepted |
| 17 | 455 | 1.2% | accepted |
| 18 | 345 | 1.2% | accepted |
| 19 | 709 | 1.1% | accepted |
| 20 | 707 | 1.1% | accepted |
| 21 | 297 | 1.1% | accepted |
| 22 | 36 | 1.1% | reject |
| 23 | 606 | 1.0% | accepted |
| 24 | 15 | 1.0% | accepted |
| 25 | 4 | 1.0% | accepted |

This data supports the proposition that woman inventors disproportionately apply for patents in classes with lower grant rates. Six of the top 25 USPC classes for female inventors have a grant rate below 50%—compared to only two of the top 25 classes for men. Moreover, for the two classes with sub-50% grant rates that appear in both genders' top-25 lists (i.e., USPC classes 705 and 424), the percentage of applications filed in these classes is higher for female inventors (4.7% for women vs 4.1% for men and 3.2% vs 1.3%, respectively). This supports our hypothesis.

Having found that female inventors work in different USPC classes and that the rejection rate for female-dominant classes is higher, we now investigate the correlation between women's rejection rates and rejected topics. To this point, we again employed an algorithm to group applications into related "clusters." Each cluster includes a group of abstracts that share a high density of similar features. The abstracts in a cluster may discuss similar technologies or use similar technical terminology, indicating that they pertain to the same or closely related technological fields. Likewise, clustered abstracts might aim at solving similar problems or build upon similar principles. To generate our clusters, we used the same approach described in Hypothesis 2.[43]

Employing the HDBSCAN algorithm, we generated a total of 1,716 subject matter clusters. Approximately 64.5% of the original dataset's abstracts were assigned to at least one cluster. Based on these results, we identified three groups of clusters (wherein only female and male clusters are mutually exclusive):

   a. **Female Clusters**: This is a cluster that disproportionately includes applications listing a woman inventor. We defined a female cluster as comprising more than 50% female applications. There are 27 (out of 1,716) clusters that meet this criterion.
   b. **Male Clusters:** This is a cluster that disproportionately includes applications listing a male inventor. We defined a male cluster as having more than 50% male applications. There are 1,682 (out of 1,716) clusters that meet this criterion. Clusters with precisely 50% males and females were considered neutral.
   c. **Reject Clusters:** This is a cluster wherein the application is disproportionately likely to be rejected. As the average rejection rate of the classes is 35% (see Figure 3), we defined a reject cluster as a cluster where more than 35% of its patent applications were rejected. Out of the 1,716 clusters, 792 "reject clusters" were identified.

Figure 3 below shows the number of clusters that have a rejection percentage within particular ranges. We see a spike in the number of clusters with a rejection rate near 30-35%, which is consistent with an average rejection rate of 35%. As we defined a "reject cluster" as being any cluster with a rejection rate over 35%, anything to the right of 35% on Figure 3's X-axis would be a reject cluster.

---

[43] *See supra* footnote 41.

**Figure 3: The Distribution of Rejection Percentage for Different Clusters**

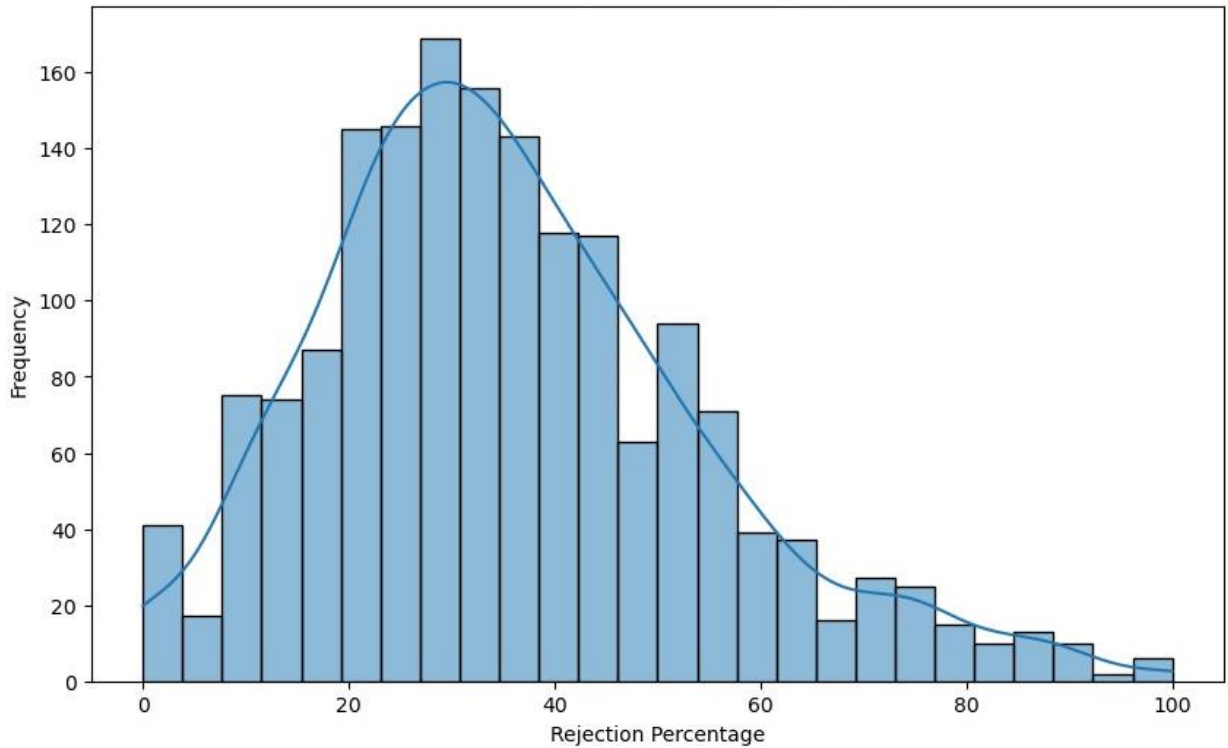

Figure 4 (below) is a Venn Diagram depicting the overlap between Male Clusters, Female Clusters, and Reject Clusters. Given the disproportionate representation of men versus women in our data, we expect to see the Venn Diagram's circle for male inventors to significantly outsize the circle for female inventors. Moreover, as we defined a Reject Cluster to have an above average rejection rate, we would expect its representative circle to embody about half of the patent applications in our study. Figure 4 supports both expectations.

The notable finding within our data is shown in the overlap of Male/Reject Clusters and Female/Reject Clusters. Over 85% (23 out of 27) of the Female Clusters are also Reject Clusters. In contrast, just over 45% (762 of 1,682) of Male Clusters are also Reject Clusters.

**Figure 4: Overlap between Male Clusters, Female Clusters, and Reject Clusters**

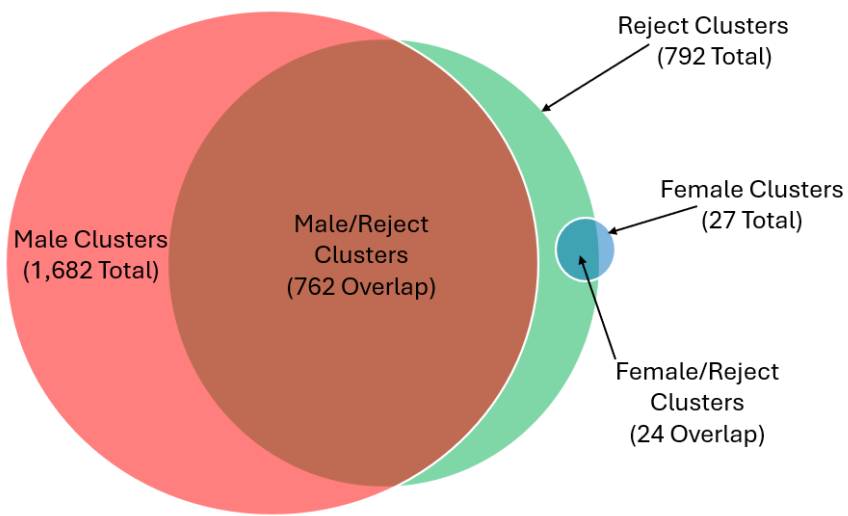

As seen in Figure 4, although women inventors make up only ~10% of the dataset, they are over-represented in clusters with high rejection rates (over 35% rejection). Specifically, 85% of female clusters are also reject clusters, while only 45% of our male clusters are also reject clusters.

Given that clusters were defined by shared keywords, we can identify the most common keywords in male and female clusters. Among male clusters, the top keywords were system, portion, device, end, member, surface, body, plurality, position, method, and support. A review of this list shows it largely consists of generic (facially non-gendered) words used in patent applications.

The list for female clusters is different. It includes terms that may—for a segment of the population—evoke feminine connotations. Common keywords from female clusters include hair, device, portion, system, child, clothing, end, surface, body, support, and breast. This connection between "feminine" words and female inventors is further analyzed in the following Section VII(A), wherein we empirically analyze this connection and discuss its legal relevance.

## VII.  Discussion and Further Investigation

Several points of inquiry arise from the previous section. We first expand upon our findings regarding inventor gender identification from application text. Beyond predicting gender, our algorithm also identified specific words that are highly indicative of inventor gender. As discussed below, we employ an additional analytical tool—corpus linguistics—to show that not only are these words gender-identifying, but they are disproportionately likely to identify female inventors. We then discuss how this may undermine a common proposal for mitigating certain gender biases.

Beyond that, recall that our earlier findings identified an intersection between textual choices and success in securing a patent. Subsections B and C propose several lines of additional research that build from this. We first discuss the need for future research into how *attorney* gender may influence the intersection of language usage, gender, and patent prosecution. Do male or female patent attorneys write differently and does this influence whether an application will be granted? The final part of this section discusses potential implications associated with our findings dealing with the overlap of female-dominated subject-matter clusters and high-rejection clusters. Specifically, we propose mechanisms that may drive this overlap and future research to test those mechanisms.

## A. Limited Impact of Anonymizing Patent Applications

To further understand our gender predicting models, we examine specific words that are highly predictive of an inventor's gender. Using tools from the corpus linguistics literature (i.e., analyzing text by comparing it to large bodies of prior language), we show that not only is application text predictive of inventor gender, but evidence supports the proposition that textual analysis disproportionately identifies *female* inventors.

This proposition is important because prior research identifies negative biases in patent examination that may arise when an inventor can be identified as a woman.[44] However, these biases are mitigated when the inventor's name does not immediately identify their gender (i.e., where a female inventor has an androgynous name). From this, scholars have proposed anonymizing inventor identity during examination.

The underlying theory is that without the inventor's name, it is impossible to identify their gender, which in turn precludes the creation of positive or negative biases (implicit or explicit) that might influence decisions in patent examination. Our analysis raises questions about whether anonymizing inventor names would mitigate such gender disparities.

Recall Hypothesis 1 which posited that we can predict inventor gender based on textual content. Consistent with that prediction, our models achieved up to 68% accuracy. This finding suggests that anonymizing inventor names may not fully eliminate biases that arise when examiners know inventors' gender. If the *text* of the application can identify gender, then anonymizing the inventor's name may not be enough to mask whether they are a man or woman.

To further investigate this issue, we studied words that our Logistic Regression Model identified as being highly predictive of inventor gender. To be clear, these terms are agnostic to the inventor's gender. Rather, their presence simply indicates that the model will probably be correct in their gender identification. The following terms appeared in the list of the top 20 most important words in at least 5 of the 9 iterations we performed:

> ('hair', 9 iterations), ('child', 9), ('decorative', 9), ('infant', 9), ('cosmetic', 9), ('garment', 9), ('jewelry', 9), ('breast', 9), ('user', 9), ('pet', 9), ('wearer', 8), ('clothing', 8), ('baby', 8), ('skin', 6), ('article', 6), ('pillow', 5), ('learning', 5), ('bra', 5), ('covering', 5), ('strap', 5)

At a glance, some of these terms, like "bra" and "cosmetic," may suggest connections to women. Given past research finding that women inventors disproportionately create inventions designed for women (as opposed to male inventors),[45] these feminine associations may suggest that not only are these words predictive of inventor gender, but they may be disproportionately predictive of *women* inventors.

Indeed, a brief analysis of the first example in that list ("hair") supports this proposition. The word "hair" occurs in over 900 abstracts in our dataset—including applications number 13/642,199 (describing a "hair band"), 13/737,943 (a "comb [that] guides the hair"), and 13/738,571 ("[a] hair curling device"). If "hair" applications were evenly distributed by gender, we would expect about 9 male applications for every female one, because our dataset has ~9 male inventor applications for every female. However, we see almost 11 times as many female "hair" applications as expected. Restated, the word "hair" probably identifies a female inventor's application.

---

[44] Jensen, et al., *supra* note 6, at 307-09; Schuster, et al., *supra* note 6, at 281-319.
[45] Rembrand Koning, Sampsa Samila & John-Paul Ferguson, *Who Do We Invent For? Patents by Women Focus More on Women's Health, but Few Women Get to Invent*, 372 SCIENCE 1345, 1345-1348 (June 18, 2021).

However, analysis of a single term and intuitive beliefs about what constitutes a "feminine" word are not sufficient to make generalized statements about gender and the predictive qualities of words in patent abstracts. To determine whether objective evidence connects these terms with female inventors, we turned to large scale textual analysis—an approach called corpus linguistics.

Corpus linguistics empirically assesses language through the study of large databases of text.[46] By analyzing how language is naturally used, researchers can draw conclusions about the meaning and usage of particular words or phrases.[47] We took the above terms (that are disproportionately likely to identify an inventor's gender in our classifier algorithm) and analyzed how they are used in natural language to learn whether they have masculine or feminine associations.

To this point, we evaluated how often the relevant terms appeared near gender-specific pronouns (i.e., his or hers) in the Corpus of Contemporary American English (COCA)—a corpus of texts from multiple genres containing over 25 million words.[48] If a term is disproportionately used next to a feminine pronoun (i.e., her bra),[49] this provides evidence that it may have feminine connotations. Applied to the current question, using a term with feminine connotations in a patent application may indicate that the inventor is a woman. This proposition is, again, premised upon prior research finding that women disproportionately create "feminine" inventions.

Using this approach, we identified 16 terms that occurred at least 100 times in the COCA corpus near a gendered pronoun.[50] Thirteen of them were disproportionately feminine: bra (occurred near "her" 7,748.0% more often than random chance would predict), jewelry (499.8%), baby (357.8%), breast (320.3%), child (208.0%), infant (207.6%), hair (118.0%), strap (86.7%), covering (45.7%), skin (43.1%), clothing (31.5%), pillow (25.0%), and learning (2.2%). Three entries in our list of gender-identifying terms were disproportionately more likely to appear in proximity to the masculine pronoun "his": pet (13.6%—indicating that "pet" occurred near "his" 13.6% more often than would have been expected given a random association to "his" or "her"), garment (76.1%), and article (95.7%).

---

[46] Richards v. Cox, 2019 UT 57, ¶ 20, 450 P.3d 1074.
[47] Eric Talbot Jensen & James Rex Lee, *International Law: Corpus Linguistics and Ordinary Meaning*, 54 GEO. WASH. INT'L L. REV. 1, 4 (2022).
[48] "The corpus contains more than one billion words of text (25+ million words each year 1990-2019) from eight genres: spoken, fiction, popular magazines, newspapers, academic texts, TV and movies subtitles, blogs, and other web pages." COCA Website Homepage
[49] To this end, we used the COCA website's "search for phrases and strings" function. To begin, we added a relevant term into a feminine or masculine string. For example, when using the word "hair" the input would be:
    1.    when appearing directly next to the modifier: "his hair"
    2.    within a single-word string adjacent to the modifier: "his * hair"
The asterisk (*) denotes any single word that could appear in between the modifier and the relevant word, for example: "his wavy hair" or "his long hair." Once we input this string, we identified the frequency of the masculine and feminine versions of these strings in COCA. To this end, we added the frequency of both observations (directly next to, and within a single-word string) to obtain the overall frequency for the male and female versions of a term.
Then, with the sum of both observations, divide the number of total "feminine" returns by the total "masculine" returns to create a ratio relative to the "feminine" modifier. Greater ratios indicate a word's greater frequency when appearing next to "her" versus "his" in COCA. Using the same method, but this time searching for only the modifiers **his** and **her** (with no terms from our above lists), we found their total appearance in COCA, which came to a ratio of ~1.30, slightly favoring **his**. This ratio could be used to identify a baseline for the occurrences of the terms "his" and "her", which is important as we would expect "his hair" to appear more often than "her hair" if the term "his" appears 1.3 times more often than "her" in general. Using this base ratio, we adjusted our observed ratios to account for the imbalance throughout COCA. Specifically, we took the base ratio and multiplied it by the previously found ratios to account for the imbalance between "his" and "her" in the data set.
[50] There were five terms that did not occur at least 100 times: decorative, cosmetic, user, wearer, and applicator.

Recognizing the disproportionate number of "female" entries in our lists (13 to 3) and the disproportionate quantum of these entries—an average of 745.7% overrepresentation[51] (near "her") for female-associated terms versus an over-representation of 61.8% (near "his") for male-centric terms—our data suggests that the above terms have disproportionately feminine connotations. Given that those terms are most likely to identify an inventor's gender, this may support the contention that it is disproportionately easy to identify a woman inventor from the text used in their patent applications. Again, the connection between "feminine" terms and women inventors is based on prior research showing that women inventors disproportionately create "feminine" inventions (i.e., inventions for women).[52]

Additional analysis reinforces this conclusion. If these terms (i.e., our list of gender indicating words) are truly "feminine," we expect women inventors to use them more often than men. That is exactly what we found. We calculated the ratio of "female" applications containing a specific word compared to "male" applications (the "W:M Word Ratio"). Then, to correct for the ~8.76:1 man:woman inventor ratio, we normalized the W:M Word Ratio by multiplying it by 8.76. One would expect this to yield normalized ratios of approximately 1 if a word is used equally by inventors of both genders.

We first tested the most common terms in the English language: "the", "be", "to", and "of." There is no reason any of these words would be gendered in usage. Accordingly, we expected them to occur in applications from men and women inventors equally often (i.e., with a W:M Word Ratio of 1). That is precisely what we found: the ratios for those terms were 1.00, 1.06, 0.99, and 1.00.

However, when we calculate this relationship for terms identified as having feminine connotations through our corpus linguistics analysis, we find substantially elevated ratios—indicating that these terms are used more frequently in patents with women inventors. Those terms (and their normalized W:M Word Ratios) are: bra (32.09), jewelry (11.73), baby (11.23), breast (8.5), child (4.89), infant (9.84), hair (8.66), strap (2.9), covering (2.25), skin (2.9), clothing (5.71), pillow (5.01), and learning (1.5). This supports the conclusion that the "feminine" words identified through our corpus linguistics analysis are disproportionately used in patent applications naming women inventors.[53]

---

[51] Even if we drop the highly disproportionate term "bra," the overrepresentation is on average 162.1%.

[52] Koning, *et al., supra* note 45.

[53] Indeed, analysis actually supports the proposition that even more of the terms identified as being highly gender identifying are feminine. Even the terms that were identified as having masculine connotations by the corpus linguistics approach were disproportionately used by female inventors in our dataset. Those terms (and ratios) were: pet (4.21), garment (7.06), and article (2.12).

Evidence does not support that this is an instance where *all* substantive words seem to be over-represented in the applications of female inventors. As suggested by ChatGPT when asked for "masculine" words, terms such as "beard", "barbell", and "grill" are all underrepresented in applications filed by women (with normalized ratios of .45, .57, and .59, respectively).

These ratios were calculated using all applications in our dataset (not just mechanical engineering and transport). However, this makes no difference to the analysis. For example, if we look only to mechanical engineering and transport applications, the first few ratios are: bra (38.12), jewelry (12.01), baby (9.78), breast (12.94). Each of these is similar in quantum to that reported in the body.

In application, these findings suggest that proposals to reduce gender biases by anonymizing review of patent applications may not be sufficient. Such proposals stand on the proposition that patent examiners identify female inventors through their names, and after gender has been identified, negative (gendered) biases decrease the chances that a woman inventor will receive a patent. But our data suggests this proposed intervention may not be enough to cure the disparity.

We find that—even without considering an inventor's name—it is possible to identify their gender over 2/3 of the time from textual choices. Furthermore, textual choices that are highly indicative of inventor gender are disproportionately likely to be highly indicative of female inventors. For example, our classifier algorithm identified the term "bra" as being particularly helpful in guessing inventor gender, but later corpus linguistics analysis shows that it is actually particularly helpful at identifying *female* inventors. These results suggest that not only can inventor gender be identified through language usage, but *women* inventors are even more likely to be identified through textual choices.

We recognize that analysis undertaken using classifier systems may not mirror human evaluators. Whether people identify inventor gender in a similar fashion is a question for future research. However, it does not seem a far reach to expect human evaluators to guess that patent applications that mention a "bra" or a "breast" are disproportionately likely to be invented by women (as suggested by our linguistic analysis).

### B. The Significance of Text Characteristics

As identified in our analysis of Hypothesis 2, textual choices can predict the probability of an application being granted. Indeed, non-substantive textual attributes of patent abstracts were the most important factors in predicting whether a patent would issue. Abstracts with shorter words are more likely to be rejected. Less readable abstracts are more likely to face rejection. Abstracts with a greater diversity of vocabulary were less likely to be granted.

These considerations raise questions about the interplay between gender and textual choices. Are those who draft patent applications for women inventors more likely to make textual choices that disproportionately hurt the odds that their applications will be granted? Are inventors more likely to hire an inventor of their gender, who in turn may influence the likelihood that they will receive a patent by engaging in a gendered writing style? Given past research finding gender differences in writing styles[54] and our research finding that patent grant rates may be influenced by writing style, substantial questions arise about whether gendered differences in patenting outcomes are driven by gendered writing styles. These questions are fertile ground for further research.[55]

### C. Potential Bias Towards Feminine Topics

The clustering analysis employed in investigating Hypothesis 3 revealed a significant overlap between fields of invention with a high number of women inventors and fields with a higher-than-average rejection rate. While the mechanisms underlying these disparities cannot immediately be ascertained from our research, several potential issues warrant future inquiry.

To begin, the possibility of a confounding variable(s) must be considered. It may be that the lower grant rate is caused by some variable that strongly correlates with women inventors and/or the fields that they invent in. For example, evidence shows that women inventors are less likely to respond to an office action rejection despite the fact that a failure to respond renders the application abandoned.[56] This disparity may be a consequence of greater risk aversion or limited access to the substantial economic resources needed to prosecute a patent and respond to USPTO office actions. Likewise, it is possible that examiners maintain some unrealized bias against particular types of technology that are disproportionately created by female inventors. In any of these cases, the fact that an invention is claimed by a female inventor is correlated to a negative outcome but would not necessarily be causal.

---

[54] *See* Stamatatos, *supra* note 14; Alluqmani, *supra* note 14.
[55] To explore the potential for improved accuracy, it would be worthwhile to train future classifiers on the basis of technological categories. By tailoring the classifier to individual categories, it may be possible to uncover nuances and factors that influence acceptance or rejection within each category.

Another potential mechanism driving disparate grant rates involves legal representation. It may be disproportionately difficult to find patent attorneys or agents willing to take seriously and credit novel improvements to products designed for women. As an example, the woman who invented Spanx undergarments wanted a woman patent attorney to prosecute her patent application because she believed it would be easier to explain her invention to a woman.[57] She discovered, however, that there were no woman patent attorneys available in her state and found that it was difficult to explain the need for her invention to a male attorney.

Such experiences, if widespread, could limit potential representation by eliminating many male patent attorneys who wouldn't prosecute an application if they do not understand the inventive contribution. It does not follow that women attorneys or agents couldn't provide effective representation; this simply means that it may be harder to find effective counsel due to a smaller set of candidates. This would make it more difficult to obtain a patent—all else being equal.

At this point, we are unable to identify any one (or all) of these potential mechanisms as driving the overlap between women inventors, "feminine" technologies, and high rejection rates. Our research does, however, illustrate the need for future study of these hypotheses to understand the correlations.

## VIII. Conclusion

The findings of this study cast new light on gendered dynamics within the patent system. Our approach—driven by text-based analysis—suggests that gender can often be discerned from the language used in patent applications. This undermines the potential efficacy of previous calls for anonymized patent examination to mitigate gendered outcomes arising from implicit or explicit biases. Indeed, our results demonstrate that women inventors, despite anonymization, could still face disparate outcomes because textual cues related to gender remain identifiable.

Additionally, our analysis finds a correlation of high rejection rates and fields with disproportionately elevated numbers of women inventors. These findings raise a host of questions relating to whether the fields in which women invent are less likely to be seen as satisfying the standards for a patent. These issues underscore the need for the patent system to further investigate how gendered attributes of invention can skew fairness in patent prosecution.

Looking forward, the study's implications extend beyond our current work. Future research should delve into the causal relationships underlying these disparities and explore mechanisms that drive gendered outcomes.

---

[56] Jordana Goodman & W. Michael Schuster, *Gender Inventorship Equity in Patent Prosecution*, __ SCIENTIFIC REPORTS __ (forthcoming) .

[57] *Spanx's Founder Couldn't Afford a Patent Attorney. So She Figured Out How to Protect Her IP Herself.*, PITCHMARK (Jan. 27, 2022, 10:34 AM), https://www.mynewsdesk.com/sg/pitchmark/news/spanxs-founder-couldnt-afford-a-patent-attorney-so-she-figured-out-to-protect-her-ip-herself-441341 [https://perma.cc/TP93-8PUR]; Jordana R. Goodman & Khamal Patterson, *Access to Justice for Black Inventors*, 77 VAND. L. REV. 109, 167 (2024).

## IX. Appendix A: Data Description and Descriptive Statistics:

Our original dataset (before cleaning and pre-processing) contained 272,401 records. Each record contains data about a patent application that was applied to the USPTO between the years 2013–20. The dataset contains the following columns:

**Table A1: Description of Included Data**

| Name | Type | Description |
|---|---|---|
| application number | Numeric | Unique sequence number for each application |
| USPC class | Mixed | USPC Patent class. There are 429 unique classes |
| one_if_male, one_if_female | Boolean | States the applicant's Gender |
| filing year | Numeric | Year the application was filed |
| one_if_patented, one_if_abandoned, one_if_pending | Boolean | Application status |
| one_if_small | Boolean | States if the application was filed by a small entity[58] |
| Biotechnology,Chem_Material_Engr,Comp_Archit, Comms,Semis,Trans,Mech_Engr | Boolean | Technological category based on the USPTO examination unit the application was assigned to |
| Abstract | Textual | Application abstract |

---

[58] *Small Entity Status, Micro Entity Status and Large Entity Status at USPTO*, NEUSTEL.COM (last visited Aug. 9, 2024), https://neustel.com/small-entity-status-micro-entity-status-large-entity-status-uspto/ (detailing the criteria for recognition as a small entity).

# X. Appendix B: Additional Descriptive Statistics and Graphics

After cleaning and pre-processing the dataset contained 255,728 records with 37 columns. The dataset contains application data from the years 2013–20. From the following visualization we can see that the number of patent abstracts in the dataset decreases in each year. The trend is similar for all applications; however until 2019, the downward trend in applications naming a woman inventor is more moderate:

**Figure B1: Percentage of Filings in Data by Gender and Filing Year**

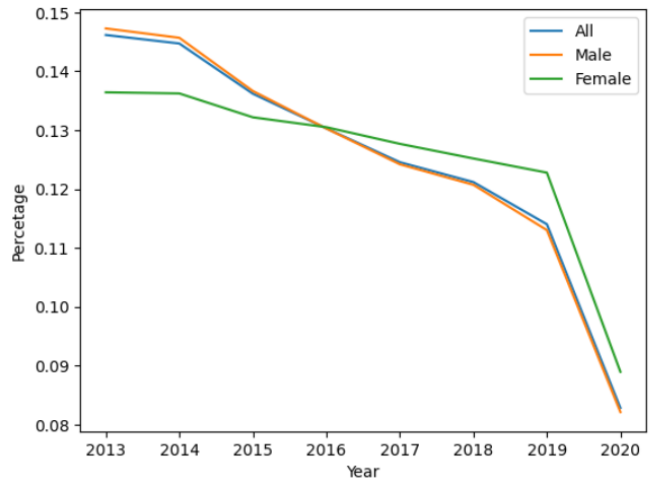

Patent applications are divided into 8 categories. The most dominant categories are Transportation and Mechanical Engineering that compose 51% of the data.

**Figure B2: Percentage of Filings by Coarse Technological Field**

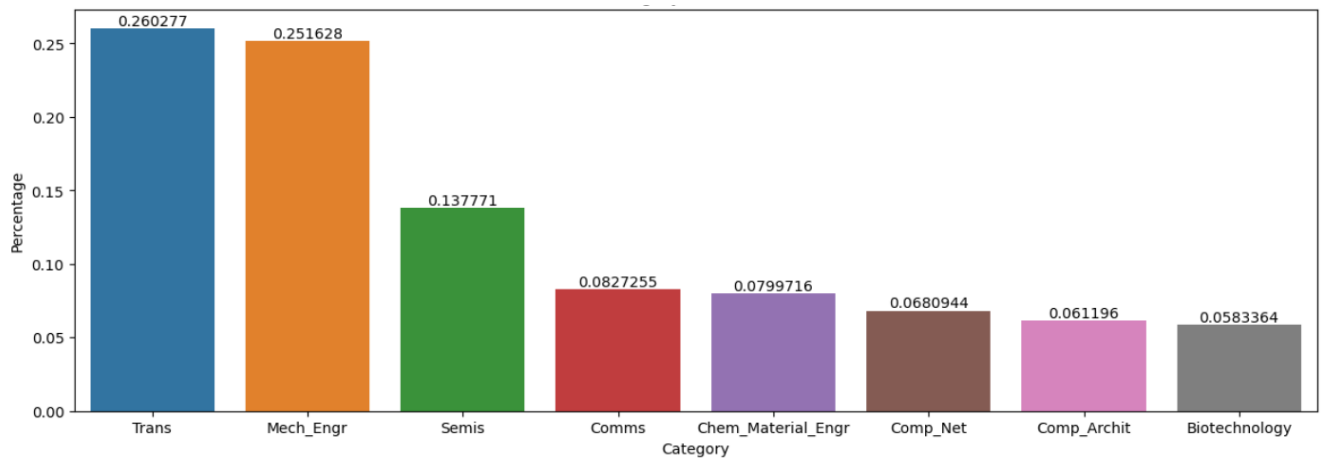

The data is highly imbalanced regarding gender. The Male-Female ratio is approximately 9 to 1.

**Figure B3: Percentage of Filings in Data by Gender**

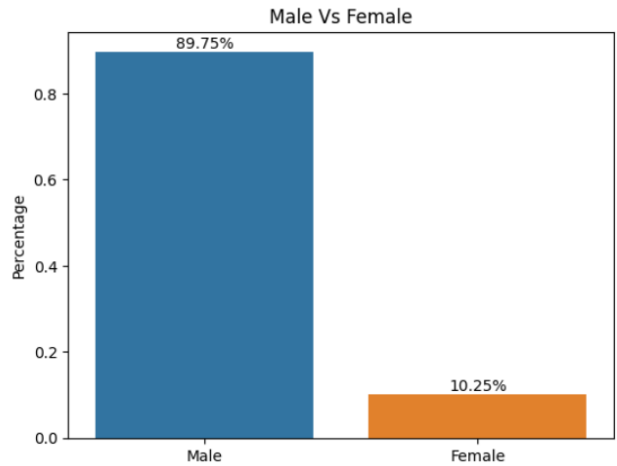

Sixty-three percent of patent applications were accepted and 27.5% were rejected.

The reject rate in female applications (36%) is much higher than the male reject rate 26%. This too is consistent with past research showing that female inventors are less likely to successfully secure a granted patent.[59]

**Figure B4: Percentage of Filings in Data by Gender and Status**

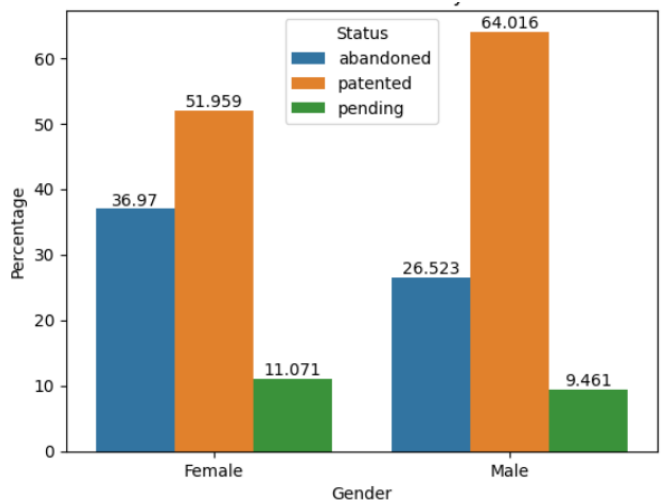

The difference in reject rate is higher for applications naming women inventors across all categories.

---

[59] See W. Michael Schuster, et. al., *supra* note 6, at 304.

**Figure B5: Percentage of Filings in Data by Gender, Coarse Technological Field, and Rejection Rate**

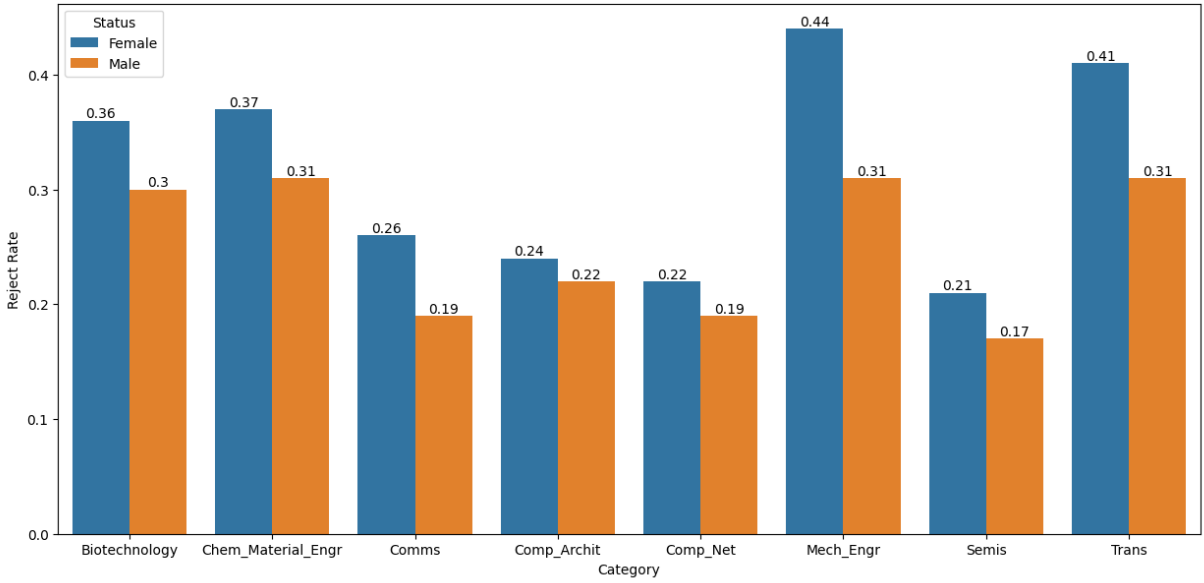

# XI. Appendix C: Gender-Based Analysis of Textual Characteristics

Figure C1 presents a full set of graphs depicting our analysis of gender-based textual characteristics. But first, we summarize the metrics and findings:

- Flesch-Kinkade ("kinkade") - This measures the Flesch Kincaid readability score. Higher values indicate lower readability. The boxplot shows that men have a slightly higher median Kincaid score compared to women.
- Word Count ("word_count") - This measures the total number of words in the application. Men have a slightly higher median word count than women, though the difference is small.
- Character Count ("char_count") - This measures the total number of characters in the text. The boxplot shows that men produce longer texts on average, with a higher median character count than women.
- Median Word Length ("median_word_length") - This measures the median length of words in the text.
- Average Word Length ("avg_word_length") - This measures the average length of words in the text. Men use slightly longer words, on average, than women.
- Skewness of Word Length ("skew_word_length") - This measures the asymmetry (skewness) in the distribution of word lengths. Women show higher skewness, suggesting a more uneven distribution of word lengths compared to men.
- Characters per Word ("characters_per_word") - This measures the average number of characters per word. Men use words with slightly more characters on average than women.
- Syllables per Word ("syll_per_word") - This measures the average number of syllables per word. The boxplot shows that men use words with slightly more syllables on average than women.
- Words per Sentence ("words_per_sentence") - This measures the average number of words per sentence. Men have a slightly higher median number of words per sentence compared to women.
- Sentences per Paragraph ("sentences_per_paragraph") - This measures the average number of sentences per paragraph.
- Type-Token Ratio ("type_token_ratio") - This measures vocabulary diversity, calculated as the ratio of unique words to total words. Women have a slightly higher type-token ratio, indicating greater lexical variety compared to men.
- Syllables ("syllables") - This measures the total number of syllables in the text. Men have a higher total syllable count, suggesting longer or more complex texts compared to women.
- Sentences ("sentences") - This measures the total number of sentences in the text.
- Long Words ("long_words") - This measures the number of long words, defined as words with six or more characters. Men use more long words on average than women.
- Complex Words ("complex_words") - This measures the number of complex words (3+ syllables). Men use more complex words on average compared to women.

# Figure C1: Textual Characteristics by Gender

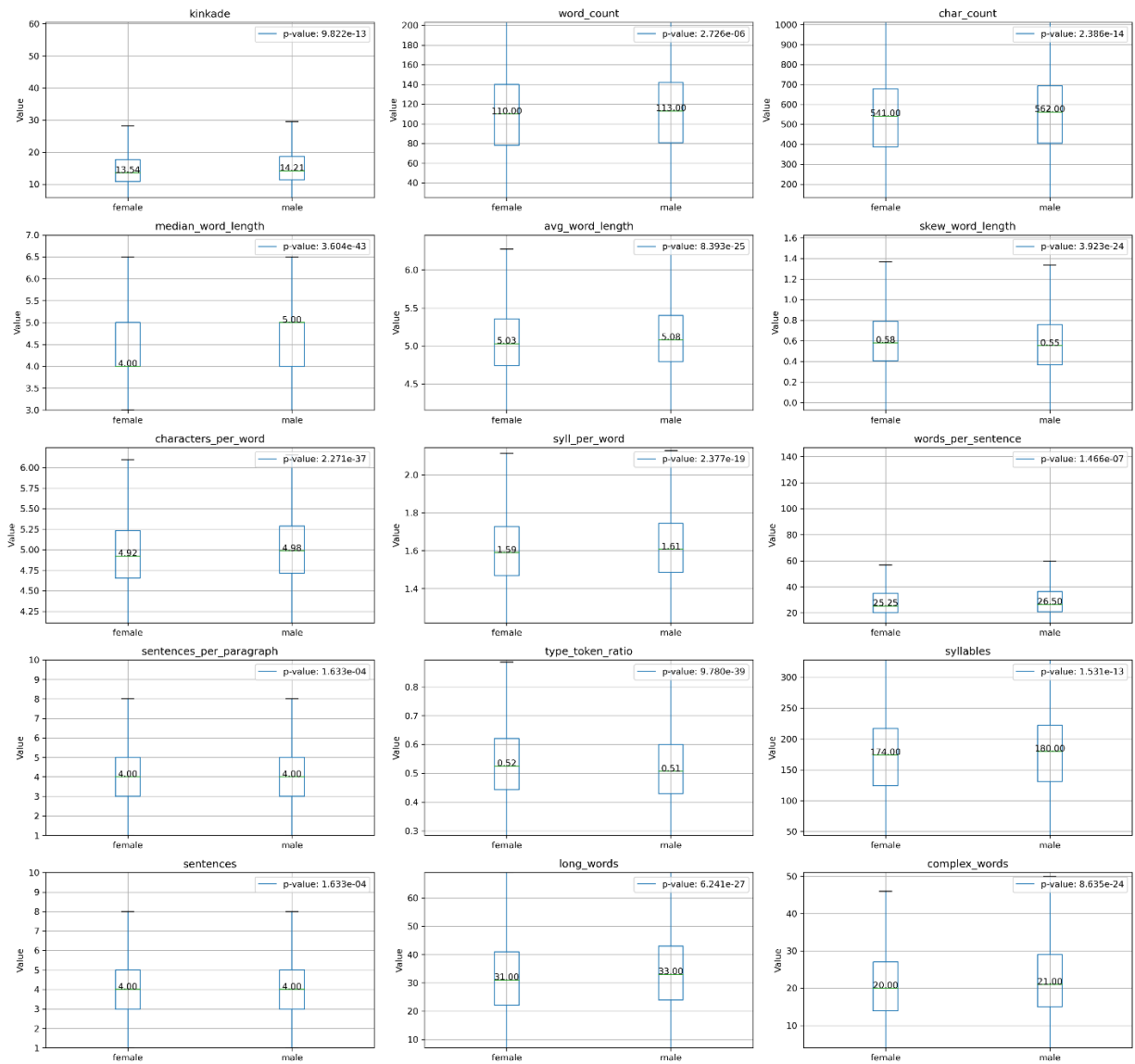

# XII. Appendix D: Application-Status Analysis of Textual Characteristics

We can additionally break down our classifier's success rates by class and inventor gender. For class, we see a slightly higher rate of correct classification for applications from mechanical engineering relative to transportation. However, this improvement is not statistically significant.

- **Mechanical Engineering Applications:**
    - **Overall Accuracy:** 68.282%
    - True Positives (correctly predicting that was granted): 34.08%
    - False Positives: 15.79%
    - True Negatives (correctly predicting that an application was rejected): 34.20%
    - False Negatives: 15.91%
- **Transportation Applications:**
    - **Overall Accuracy:** 64.814%
    - True Positives: 32.01%
    - False Positives: 17.19%
    - True Negatives: 32.80%
    - False Negatives: 17.98%

The classifier likewise attains higher accuracy for applications naming female inventors (64.81%), relative to those with a male inventor (60.93%). For the Female category, the True Positive rate (40.87%) is quite high relative to the other categories, but there is also a high False Negative rate (34.04%).

- **Applications from a Male Inventor:**
    - **Overall Accuracy:** 60.93%
    - True Positives: 26.27%
    - False Positives: 23.72%
    - True Negatives: 34.66%
    - False Negatives: 15.33%
- **Applications from a Female Inventor:**
    - **Overall Accuracy:** 64.814%
    - True Positives: 40.87%
    - False Positives: 9.12%
    - True Negatives: 15.95%
    - False Negatives: 34.04%

Additionally, to assess the influence of different textual features on each patent class (granted and rejected), we examined their distribution within each category. Attributes of granted patents are shown on the left and abandoned applications on the right. Each rectangular box represents the interquartile range—the range between the first quartile (25th percentile) and the third quartile (75th percentile). This means that 50% of the data points fall within this range. The height of the box gives a visual indication of the range of the data and the horizontal line within the rectangle represents the median of the dataset. The horizontal lines (whiskers) above and below the rectangles represent the highest and lowest values within the dataset, excluding outliers which are represented as the dots above or below the whiskers.

**Figure D1: Textual Characteristics by Grant Status**

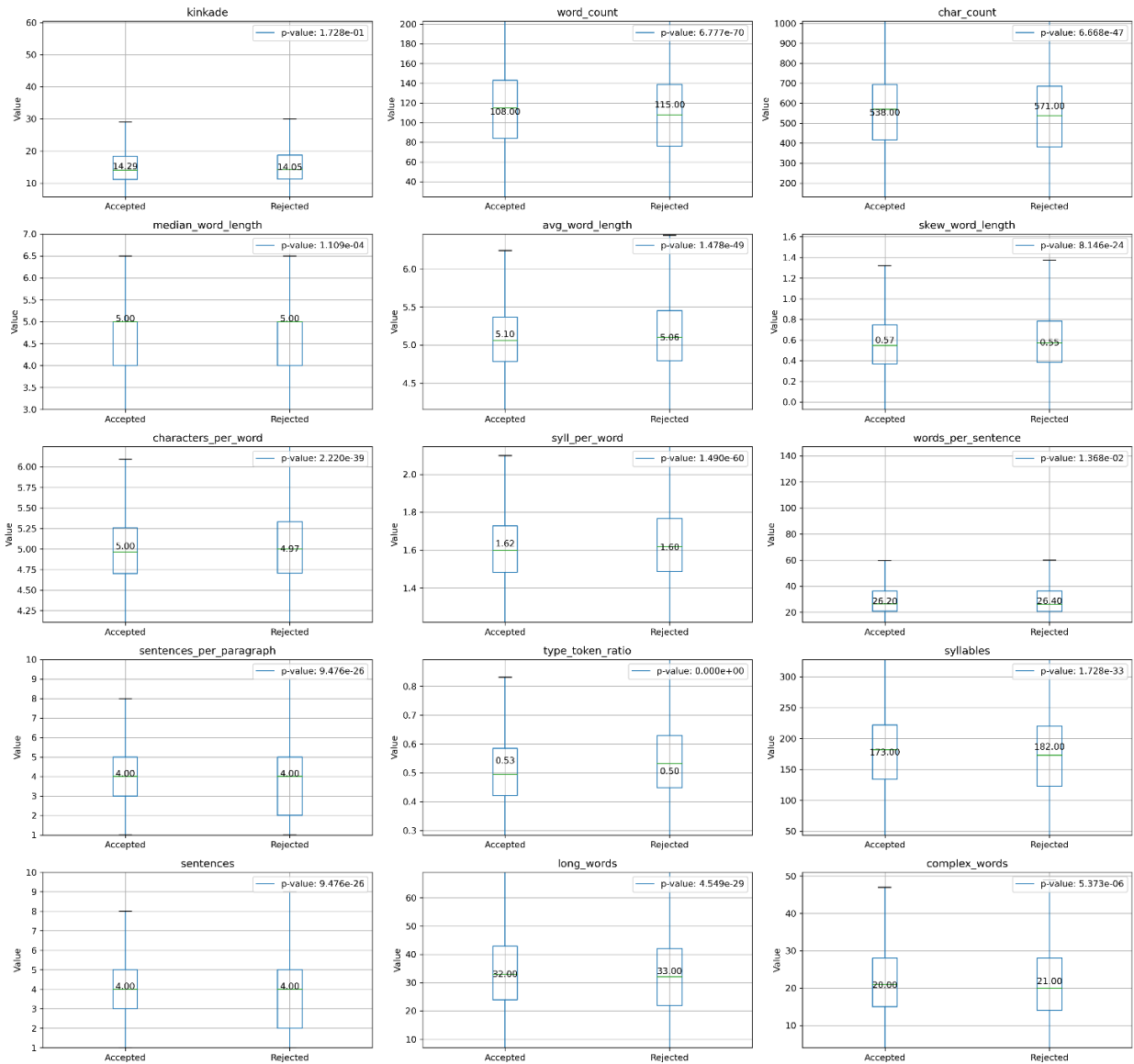

As an example, we consider type-token ratio to be the most important feature in predicting patent grant, as indicated by our classifier. The left-most part of the graphic deals with the type-token ratio for granted and rejected. Type-token ratio measures the variety of words used in the text. Lower values may indicate repetitive use of the same terms, while higher values suggest the use of a broader vocabulary. Here, we see the rejected class box, its median, and associated whiskers are all higher than the accepted class box. This indicates that, on average, the type-token ratio for rejected applications is greater—meaning that rejected applications tend to use a broader range of words in their abstract.